\documentclass[epj]{webofc}
\usepackage[varg]{txfonts}   
\usepackage{graphicx,color} 
\usepackage{bm}       
\usepackage{amsmath}  
\usepackage{amssymb}
\usepackage{epstopdf}
\woctitle{21st International Conference on Few-Body Problems in Physics}
\begin{document}
\title{Quarkonia and heavy-light mesons in a covariant quark model}
\author{Sofia Leit\~ao\inst{1}\fnsep\thanks{\email{sofia.leitao@tecnico.ulisboa.pt}} \and
        Alfred Stadler\inst{2,1}\fnsep\thanks{\email{stadler@uevora.pt}} \and
        M. T. Pe\~na\inst{1,3}\fnsep\thanks{\email{teresa.pena@tecnico.ulisboa.pt}} \and
        Elmar P. Biernat\inst{1}\fnsep\thanks{\email{elmar.biernat@tecnico.ulisboa.pt}}
}

\institute{Centro de F\'isica Te\'orica de Part\'iculas, Instituto Superior T\'ecnico (IST), Universidade de Lisboa, 1049-001 Lisboa, Portugal 
\and
           Departamento de F\'isica, Universidade de \'Evora, 7000-671 \'Evora, Portugal 
\and
           Departamento de F\'isica, Instituto Superior T\'ecnico (IST), Universidade de Lisboa, 1049-001 Lisboa, Portugal
          }

\abstract{
  Preliminary calculations using the Covariant Spectator Theory (CST) employed a scalar linear confining interaction and an additional constant vector potential to compute the mesonic mass spectra. In this work we generalize the confining interaction to include more general structures, in particular a vector and also a pseudoscalar part, as suggested by a recent study \cite{piPi}. A one-gluon-exchange kernel is also implemented to describe the short-range part of the interaction. We solve the simplest CST approximation to the complete Bethe-Salpeter equation, the one-channel spectator equation, using a numerical technique that eliminates all singularities from the kernel. The parameters of the model are determined through a fit to the experimental pseudoscalar meson spectra, with a good agreement for both quarkonia and heavy-light states.
}
\maketitle
\section{Introduction}
\label{intro}
When applied to the description of hadrons, QCD is a highly complex nonperturbative theory that cannot be solved directly. Numerical simulations of a space-time discretized version of QCD, lattice QCD \cite{RefB}, are promising, but still have limitations. Continuum methods using the Bethe-Salpeter/Dyson-Schwinger equations \cite{Ref1, Ref2, Ref3} leave out some possibly important features of QCD, and so far do not describe all decay properties of the meson spectrum and in particular of the heavy-light sector. Moreover, both lattice QCD and Dyson-Schwinger calculations are performed in Euclidean and not in the physical Minkowski space. This simplification allows the calculation of mass spectra, but it is not clear how the obtained amplitudes can be reliably related to the physical ones. 

This work is inserted in a larger project which proposes to implement the following:
1) a calculation of quark-antiquark bound states in Minkowski space; 2) dynamical quark mass generation, due to the self-interaction of quarks calculated consistently from the quark-quark interaction kernel and satisfying chiral symmetry constraints; 3) a confining term in the interaction kernel that in the nonrelativistic limit reduces to a linear potential.
The parameters of the interaction kernel will be determined by fits of the meson spectra, and are constrained by the quark mass functions obtained in lattice QCD. This will also provide information on the Lorentz structure of the confining interaction, which is not well known at present.\ The dressed electromagnetic quark current will be calculated consistently, using the same kernel. 

Once the quark current is known, many more applications become possible, such as hadron form factors and decay properties. We single out the calculation of hadronic contributions to light-by-light scattering, which is the most uncertain part of the theoretical prediction of the muon’s anomalous magnetic moment. This uncertainty needs to be reduced in order to find out if the observed 3$\sigma$ discrepancy between theory and measurement is a sign of new physics beyond the Standard Model \cite{pion}.

\section{Pseudoscalar meson spectra with the one-channel CST equation}
\label{sec-1}

We use the Covariant Spectator Theory (CST) \cite{FG}, which can be viewed as a reorganization of the complete Bethe-Salpeter equation (BSE), to develop a dynamical quark model in Minkowski space that can describe the structure and the mass spectrum of both heavy and light quark systems.

\begin{figure} 
\centering
\includegraphics[width=6cm]{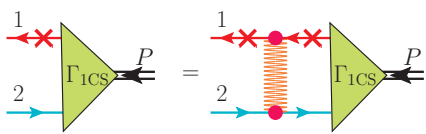}
\caption{Diagrammatic representation of the 1CSE.\ The cross indicates that particle 1 is on its mass-shell. $P$ is the total momentum and $\Gamma_\mathrm{1CS}$ the vertex function of the bound-state.} \label{fig:1CS1}
\end{figure}

The simplest version of the CST equation, called the one-channel CST equation (1CSE) and shown diagrammatically in Fig.\ \ref{fig:1CS1}, is obtained when only the positive-energy pole contribution from the heavier quark propagator in the energy loop integration is kept.

An important property of the 1CSE is that, unlike the BSE in ladder approximation, it has a smooth nonrelativistic limit, and therefore in calculations of heavy quarkonia its results can be compared directly to the ones obtained with the Schr\"odinger equation (SE).

To find an appropriate way to treat a linear confining interaction in momentum space, and to test the required numerical methods, in \cite{LC} we used the SE to calculate the two-body bound states of a linear potential. Since in $S$ waves these solutions are known exactly, this is an ideal test case. It turned out that the singularities arising in the kernel can be treated efficiently, yielding accurate and stable numerical solutions.  In this work, the same methods were applied for the first time to solve the relativistic 1CSE.

The kernel employed in our calculations with the 1CSE  consists of a covariant generalization of the linear (L) confining potential used in \cite{LC}, a one-gluon exchange (OGE), and a constant (C) interaction,
\begin{equation}
\mathcal{V}(p,k) = \mathcal{V}_\mathrm{L}(p,k) + \mathcal{V}_\mathrm{OGE}(p,k)+\mathcal{V}_\mathrm{C}(p,k) \, ,
\end{equation}
where 
\begin{align}
& \mathcal{V}_\mathrm{L}(p,k)   = -8\sigma \pi\left[\frac{1}{q^4}-\frac{E_{p_1}}{m_1}(2\pi)^3 \delta^3 (\mathbf{q})\int \frac{d^3 k'}{(2\pi)^3}\frac{m_1}{E_{k'_1}}\frac{1}{q'^4}\right] \left[ (1-y)({\bf 1} \otimes {\bf 1} + \gamma^5 \otimes \gamma^5 ) - y (\gamma^\mu \otimes \gamma_\mu)\right] \, ,
\nonumber\\
&\mathcal{V}_\mathrm{OGE}(p,k)  = -\frac{4 \pi \alpha}{q^2}( \gamma^\mu \otimes \gamma_\mu)  \, , 
\qquad \qquad
\mathcal{V}_\mathrm{C}(p,k) = (2\pi)^3\frac{E_{k_1}}{m_1} C \delta^3 (\mathbf{q})( \gamma^\mu \otimes \gamma_\mu) \, ,
\label{eq:V}
\end{align}
$q=p-k$ is the transferred four-momentum, and $E_{p_1}=(m_1^2+\vec p^2)^{1/2}$ is the energy of the heavier quark with mass $m_1$. The mixing parameter $y$  allows to dial continuously between a scalar-plus-pseudoscalar structure, suggested recently in \cite{piPi} due to chiral-symmetry constraints, and a vector structure, while preserving the same nonrelativistic limit. The precise Lorentz structure of the confining interaction is not known, and by fitting the $y$ parameter from the mesonic spectra some further information can be gained. 
The three coupling strengths $\sigma$, $\alpha$ and $C$, were treated as free model parameters.

For this preliminary work, we used constant masses for the constituent quarks, with the values
\begin{equation}
 m_b=4.793\mbox{ GeV}, \quad  m_c=1.530\mbox{ GeV}, \quad m_s=0.400\mbox{ GeV}, \quad  \text{and} \quad m_u=m_d=0.258 \mbox{ GeV} \, .
\end{equation}
The 1CSE---with retardation included---requires regularization. We used a Pauli-Villars regularization with a cut-off parameter proportional to the average mass of the two constituent quarks, $\bar{m}$: $\Lambda=1.7 \bar{m}$.

\begin{figure} 
\centering
\includegraphics[width=12.50cm]{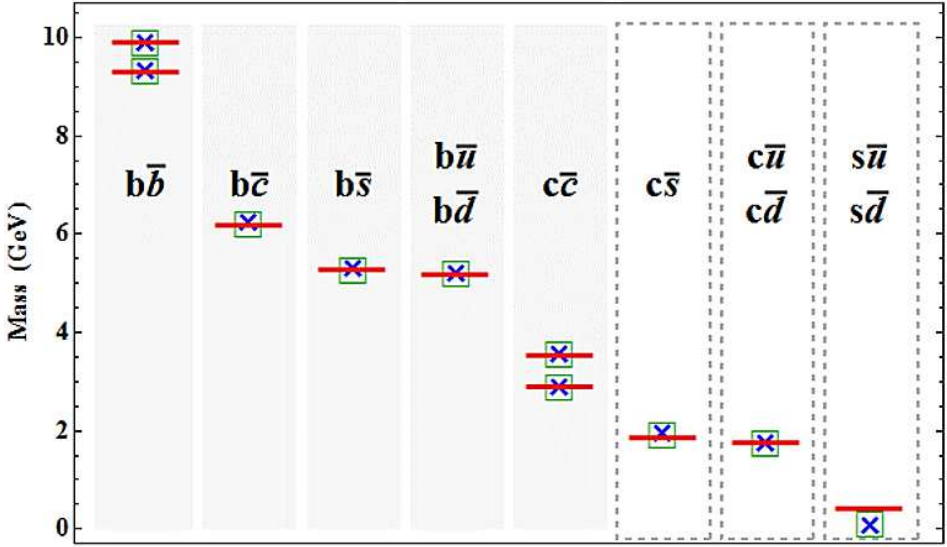}
\caption{Mass spectra computed with the 1CSE for the $b\bar{b}$, $c\bar{c}$ quarkonia, and for the heavy-light $b\bar{s}$, $b\bar{u}/b\bar{d}$, $b\bar{c}$, $c\bar{u}/c\bar{d}$, $s\bar{u}/s\bar{d}$ pseudoscalar states. The horizontal bars represent the experimental $J^{PC}=0^{--}$ states, the centers of the crosses and open squares indicate the calculated states for models 1 and 2, respectively.  The states fitted to the data are those inside the shaded regions, whereas the remaining states inside the dashed rectangles are predictions.} \label{fig:hl1}
\end{figure}

The pseudoscalar meson spectra were fitted with two distinct models with different Lorentz structure of the confining interaction: 
whereas model 2 has exactly the equally weighted scalar-pseudoscalar component, mixed with a vector structure, as given in Eq.~(\ref{eq:V}), model 1 omits the pseudoscalar part.
The obtained results are given in  Figure \ref{fig:hl1} and Table  \ref{tab-1}.

\begin{table}
\centering
\caption{Fitted parameters for kernel models 1 and 2.}
\label{tab-1}       
\begin{tabular}{|l|llll|}
\hline
 & $\sigma$ [GeV$^2$] & $\alpha$ & $C$ [GeV]& $y$\\\hline
Model 1 & $0.22$ & 0.38 & 0.337  & $8.60 \times 10^{-7}$  \\
Model 2 & $0.21$ & 0.37 & 0.311 & $0.38 \times 10^{-7}$ \\\hline
\end{tabular}
\end{table}

We find that good fits of the pseudoscalar meson spectra can be obtained with either model, and that the predictions for the mesons containing lighter quarks are also very reasonable. The observed slightly larger difference between the model predictions and the data for the lightest systems is in part due to a pole in the propagator of the lighter quark whose importance increases with decreasing bound-state mass, but which is omitted in the 1CSE. This pole will be included in future work.

It turns out that both model fits favor essentially pure Lorentz scalar structures of the confining interaction. It will be interesting to see if this feature persists once the fits are extended to include also vector mesons.

The inclusion of a pseudoscalar part in the confining interaction leads to a marginally better fit. However, because the heavier mesons are almost nonrelativistic systems for which the pseudoscalar interaction is strongly suppressed, the difference between models 1 and 2 becomes noticable only for the lighter states. In these cases, on the other hand, it may become necessary to replace the 1CSE by a more complete version of the CST bound-state equations, namely the two-channel  (2CSE) or---in the case of the pion---the complete  four-channel spectator equation (4CSE) \cite{MF}.

\section{Conclusions and Outlook}
The numerical techniques developed so far allowed us to solve the 1CSE and perform the first fit with a realistic linear-plus-Coulomb-plus-constant kernel to a wider range of quarkonia and heavy-light pseudoscalar states.

As an immediate continuation of this work, we plan to extend the fit to other mesons (most importantly vector mesons), and to investigate the effect of replacing the constant constituent quark masses by quark mass functions that are calculated in a fully self-consistent fashion from the quark-antiquark kernel. The next goal will be to solve the 2CSE and 4CSE to obtain reliable solutions also for the light sector.

Meanwhile, based on the results determined so far, we can already state that the calculational methods we developed to treat a  linear confining interaction in momentum space and tested in the nonrelativistic SE, also work in the relativistic case of the 1CSE in Minkowski space. These techniques will be useful in future studies of the mesonic bound-state problem using the CST.

\begin{acknowledgement}
This work received financial support from Funda\c c\~ao para a Ci\^encia e Tecnologia (FCT) under Grant No.\ CFTP-FCT (PEst-OE/FIS/U/0777/2013). 
\end{acknowledgement}

%
%
%

\end{document}